\documentclass[prd,aps,eqsecnum,floatfix,nofootinbib,preprint,tightenlines]{revtex4}

\usepackage{latexsym}
\usepackage{graphicx}
\usepackage{amsmath}

\usepackage[pdftex]{color}
\usepackage[colorlinks=true,linkcolor=blue,filecolor=blue,urlcolor=blue,citecolor=blue,pdftex=true,plainpages=false]{hyperref}

\def\bibi{\bibitem}
\def\floatcaption#1#2{ \caption{#2 \label{#1}} }

\let\ced=\c                     

\let\inodot=\i

\def\a{\alpha}
\def\b{\beta}
\def\c{\chi}
\def\d{\delta}
\def\g{\gamma}

\def\i{\iota}

\def\m{\mu}

\def\p{\pi}                     
\def\r{\rho}                    
\def\t{\tau}

\def\D{\Delta}

\def\P{\Pi}

\def\cbo{{\,\raise-.15ex\Sc [\,}}                       

\def\ddt#1{{\buildrel {\hbox{\LARGE .\kern-2pt.}} \over {#1}}}

\def\ie{\mbox{\it i.e.}}

\def\etc{\mbox{\it etc.}}

\long\def\symbolfootnote[#1]#2{\begingroup%
\def\thefootnote{\fnsymbol{footnote}}\footnote[#1]{#2}\endgroup}

\long \def \blockcomment #1\endcomment{}

\def\seef{{\it cf.}}

\begin{document}
\rightline{TUM-HEP-872/12}
\rightline{UAB-FT-728}

\thispagestyle{empty}

\begin{center}
\vspace*{5mm}
\begin{boldmath}
{\large\bf Low-energy constants and condensates from the  $\tau$ hadronic spectral functions}
\end{boldmath}
\\[10mm]
Diogo Boito,$^a$
Maarten Golterman,$^b$\symbolfootnote[2]{Permanent address: Department of Physics and Astronomy,
San Francisco State University, San Francisco, CA 94132, USA}
Matthias Jamin,$^c$ Kim Maltman,$^{d,e}$\\ Santiago Peris$^f$
\\[8mm]
{\small\it
$^a$Physik Department T31, Technische Universit\"at M\"unchen,
James-Franck-Stra\ss e 1\\ D-85748 Garching, Germany
\\[5mm]
$^b$Institut de F\'\inodot sica d'Altes Energies (IFAE), 
Universitat Aut\`onoma de Barcelona\\ E-08193 Bellaterra, Barcelona, Spain
\\[5mm]
$^c$Instituci\'o Catalana de Recerca i Estudis Avan\ced{c}ats (ICREA),
IFAE\\  Universitat Aut\`onoma de Barcelona, E-08193 Bellaterra,
Barcelona, Spain
\\[5mm]
$^d$Department of Mathematics and Statistics,
York University\\  Toronto, ON Canada M3J~1P3
\\[5mm]
$^e$CSSM, University of Adelaide, Adelaide, SA~5005 Australia
\\[5mm]
$^f$Department of Physics, Universitat Aut\`onoma de Barcelona\\ E-08193 Bellaterra, Barcelona, Spain}
\\[10mm]
{ABSTRACT}
\\[2mm]
\end{center}

\begin{quotation}
We use results of fits to the OPAL spectral data, obtained from
non-strange hadronic $\t$ decays, to evaluate the difference
between the vector and axial current correlators, $\P_{V-A}(Q^2)$.
The behavior of $\P_{V-A}(Q^2)$ near euclidean
momentum $Q^2=0$ is used to determine the effective low-energy constants
$L_{10}^{\rm eff}$ and $C_{\rm 87}^{\rm eff}$ related to the renormalized
low-energy constants $L_{10}^r$ and $C_{87}^r$ in the chiral lagrangian.
We also investigate how well two-loop
chiral perturbation theory describes $\P_{V-A}(Q^2)$ as a function of $Q^2$. 
This is the
first determination of $L_{10}^{\rm eff}$ and $C_{87}^{\rm eff}$ to employ
a
fully self-consistent model for the violations of quark-hadron duality in 
both the vector and axial channels.
We also discuss the
values of the coefficients $C_{6,V-A}$ and $C_{8,V-A}$ governing
the dimension six and eight contributions to the operator
product expansion representation
of $\Pi_{V-A}(Q^2)$.
\end{quotation}

\newpage
\section{\label{introduction} Introduction}
Recently, we reanalyzed the OPAL spectral function data for non-strange
hadronic $\t$ decays \cite{OPAL}, the main aim being a determination of a value for the
strong coupling at the $\t$ mass, $\a_s(m_\t^2)$, with a complete error
analysis \cite{US2,US1}.   Among the new elements in this analysis were
the use of spectral-function moments with a good perturbative behavior
\cite{BBJ}, and a complete and self-consistent treatment of non-perturbative
effects \cite{US1,MY}. 
This, in turn,
requires a quantitative treatment of quark-hadron duality violations (DV) due
to the clear presence of hadronic resonances in the spectral function data.
The latter was accomplished by employing a model developed in 
Refs.~\cite{CGP05,CGP08} that we will also use in the present article.   While
the analysis necessarily relies on this model, we demonstrated that the 
complete theoretical parametrization of the spectral-function moments
including the DV part
provides a very good description of the experimental data.   We chose to
use OPAL data, rather than ALEPH data \cite{ALEPH} because of the 
incompleteness of the data correlations \cite{TAU10} for the latter.

While the central results in Refs.~\cite{US2,US1} were based on fits to only the
vector channel data, we also carried out simultaneous fits to the vector and axial
channel data as a consistency check on our results.   As a by-product, we thus
have a quantitative theoretical description of the vector and axial spectral
functions $\r_V(t)$ and $\r_A(t)$ from $t=t_{min}\approx 1.3$~GeV$^2$ to $t=\infty$.   This lets us evaluate dispersive integrals over
$\r_V(t)-\r_A(t)$ as a function of euclidean momentum $Q$ quantitatively from the data.  (For explicit expressions, see Eqs.~(\ref{defPiLR}) and~(\ref{defPiLRw}) below.)
This, in turn, allows us to extract certain low-energy constants (LECs) appearing in
the chiral lagrangian, as well as some of the coefficients appearing in the
operator product expansion (OPE), from the low and high $Q^2$ behavior of $\P_{V-A}(Q^2)$ , respectively.
The determination of these LECs and OPE coefficients is the aim of the
present article.   As we will explain in detail below, we determine 
$\P_{V-A}(Q^2)$ by summing over experimental data up to $t=t_{\rm switch}$,
and using our fitted spectral functions for $t\in[t_{\rm switch},\infty)$,
where we will choose $t_{\rm switch}\in[t_{min},m_\t^2]$ ($m_\t$ is the
$\t$ mass).

This article is organized as follows.
In Sec.~\ref{theory} we give a brief overview of the necessary theory, including
a rederivation of the Weinberg sum rules beyond the chiral limit
tailored to our analysis.   In Sec.~\ref{data} we explain our strategy for the numerical evaluation of
$\P_{V-A}(Q^2)$ and other related functions from the OPAL data.   In Sec.~\ref{results} we present and discuss our results.   We include an investigation of a fit
of $\P_{V-A}(Q^2)$ to chiral perturbation theory (ChPT) to two-loop order.
Our conclusions are contained in Sec.~\ref{conclusion}.

\section{\label{theory} Overview of theory}
The LECs and OPE condensates this article aims to extract are all related to
$\P_{V-A}(Q^2)$ defined by\footnote{For our conventions, see Ref.~\cite{US1}.}
\begin{equation}
\label{defPiLR}
\P_{V-A}(Q^2)=\int_0^\infty dt\;\frac{\r_V(t)-\r_A(t)}{t+Q^2}\ ,
\end{equation}
with $Q^2$ the euclidean external momentum, and $\r_V$ ($\r_A$) the
non-strange $I=1$ vector (axial) spectral functions summing the angular momentum $J=1$ and $J=0$
contributions.
Here and in what follows we
take, for convenience, $\r_A$ to be the axial spectral function
without the contribution from the pion pole 

The difference $\r_V-\r_A$ is constrained by the
Weinberg sum rules \cite{WSR}.   It is useful to briefly review their derivation,
beginning with the second sum rule, because of the subtleties involved at non-zero quark mass.   Following Ref.~\cite{FNR}, and showing contributions from the pion pole explicitly
because it is not contained in $\r_A(t)$, we write
\begin{eqnarray}
\label{cauchy}
\int_0^{s_0}dt\;w(t)\left(\r_V(t)-\r_A(t)\right)-2f_\p^2 w(m_\p^2)&=&-\frac{1}{2\p i}\oint_{|z|=s_0}
dz\;w(z)\;\P_{V-A}(z)\\
&&\hspace{-4.2cm}=-\frac{1}{2\p i}\oint_{|z|=s_0}
dz\;w(z)\;\P_{V-A}^{\rm OPE}(z)-\frac{1}{2\p i}\oint_{|z|=s_0}
dz\;w(z)\;\P_{V-A}^{\rm DV}(z)\ ,\nonumber
\end{eqnarray}
where $w(t)$ is a polynomial in $t$, and
where we split
\begin{equation}
\label{split}
\P_{V-A}(z)=\P_{V-A}^{\rm OPE}(z)+\P_{V-A}^{\rm DV}(z)
\end{equation}
into the OPE and duality-violating (DV) parts, following Ref.~\cite{CGP05}.
The OPE part has the form
\begin{equation}
\label{highQ2}
\P_{V-A}^{\rm OPE}(-Q^2)=\sum_{k=1}^\infty\;\frac{C_{2k,V-A}}{(Q^2)^k}\ ,
\end{equation}
with, for three flavors \cite{FNR,NLOC4}, 
\begin{subequations}
\label{C2C4}
\begin{eqnarray}
\hspace{-0.2cm}C_{2,V-A}
&\!\!\!=&\!\!\!-\frac{\a_s(\m^2)}{\p^3}\,m_u(\m^2)m_d(\m^2)\left(1-\frac{\a_s(\m^2)}{\p}\left(\frac{17}{4}\log{\frac{Q^2}{\m^2}}+c\right)\right)+\dots,\label{C2C4a}\\
\hspace{-0.2cm}C_{4,V-A}&\!\!\!=&\!\!\!-\frac{8}{3}\frac{\a_s}{\p}f_\p^2 m_\p^2+\dots\ ,\label{C2C4b}
\end{eqnarray}
\end{subequations}
where $\m$ is the renormalization scale,
$m_{u,d}(\m^2)$ denote the running up and down quark masses 
and $c$ is a numerical constant whose value
is not required in what follows. In Eq.~(\ref{C2C4b}), isospin symmetry
has been assumed, and the Gell-Mann--Oakes--Renner relation has been used to express
the product of the average light quark mass and quark condensate in
terms of $f_\p$ and $m_\p$.  Contributions from higher-dimensional
operators will be neglected.
Next, in order to derive the second Weinberg sum rule, we choose
$w(t)=t$.
Expressing the DV part of Eq.~(\ref{cauchy}) in terms of the
DV parts of the vector and axial spectral functions \cite{CGP05}, 
\begin{equation}
\label{rVrA}
\r_V^{\rm DV}(t)-\r_A^{\rm DV}(t)=\frac{1}{\p}\,\mbox{Im}\,\P^{\rm DV}_{V-A}(t)\ ,
\end{equation}
and
evaluating the OPE part using Eq.~(\ref{C2C4}), Eq.~(\ref{cauchy}) can be
rewritten as
\begin{eqnarray}
\label{WSR2}
&&\hspace{-0.5cm}\int_0^{s_0}dt\;t\left(\r_V(t)-\r_A(t)\right)+\int_{s_0}^\infty dt\;t\left(\r_V^{\rm DV}(t)-\r_A^{\rm DV}(t)\right)\\
&&\hspace{2.5cm}=2f_\p^2 m_\p^2\left(1+\frac{4}{3}\frac{\a_s(s_0)}{\p}\right)+\frac{17}{4\p^2}\left(\frac{\a_s(s_0)}{\p}\right)^2m_u(s_0)m_d(s_0)s_0
\ ,\nonumber
\end{eqnarray}
where we set $\m^2=s_0$.
This is the version of the second Weinberg sum rule we will employ.
A similar derivation, choosing
$w(t)=1$, leads to the first Weinberg sum rule,
\begin{equation}
\label{WSR1}
\int_0^{s_0} dt\;\left(\r_V(t)-\r_A(t)\right)
+\int_{s_0}^\infty dt\;\left(\r_V^{\rm DV}(t)-\r_A^{\rm DV}(t)\right)=2f_\p^2\ ,
\end{equation}
where we already dropped the correction coming from the
OPE contributions to the right-hand side of
Eq.~(\ref{cauchy}), as these are numerically tiny for the $s_0$
of interest to us.
Our conventions are such that $f_\p=92.21(14)$~MeV.

The effective LECs $L_{10}^{\rm eff}$ and $C_{87}^{\rm eff}$ are defined from the expansion of $\P_{V-A}(Q^2)$ around $Q^2=0$ \cite{DGHS,ABT,GPP}:
\begin{equation}
\label{lowQ2}
\P_{V-A}(Q^2)=-8L_{10}^{\rm eff}-16C_{87}^{\rm eff}Q^2+O(Q^4)\ ,
\end{equation}
while the OPE condensates $C_{6,V-A}$ and $C_{8,V-A}$ are defined from the high-$Q^2$ expansion~(\ref{highQ2}).

We will also use functions
$\P_{V-A}^{(w)}$ involving
additional polynomial weight factors $w(x)$, defined by
\begin{equation}
\label{defPiLRw}
\P_{V-A}^{(w)}(Q^2)=\int_0^\infty dt\;w(t/s_0)\;\frac{\r_V(t)-\r_A(t)}{t+Q^2}\ .
\end{equation}
The weights we will consider are
\begin{equation}
\label{weights}
w_k(x)=(1-x)^k\ ,\quad k=1,2\ .
\end{equation}
Using the Weinberg sum rules Eqs.~(\ref{WSR2}) and~(\ref{WSR1}), one finds
\begin{eqnarray}
\label{alt}
&&\hspace{-0.4cm}-8L_{10}^{\rm eff}=\P_{V-A}(0)=\P_{V-A}^{(w_1)}(0)+\frac{2f_\p^2}{s_0}\\
&&\hspace{0.2cm}=\P_{V-A}^{(w_2)}(0)+\frac{4f_\p^2}{s_0}\left[1-\frac{17}{16\p^2}\left(\frac{\a_s(s_0)}{\p}\right)^2\frac{m_u(s_0)m_d(s_0)}{f_\p^2}
-\frac{m_\p^2}{2s_0}\left(1+\frac{4}{3}\frac{\a_s(s_0)}{\p}\right)
\right]\ ,
\nonumber
\end{eqnarray}
yielding alternative ways to evaluate $L_{10}^{\rm eff}$.  Similar equations
can be derived for $C_{87}^{\rm eff}$.
In these equations, we assumed that $\P_{V-A}^{(w)}(Q^2)$ can be
written as in Eq.~(\ref{defPiLRw}), using the experimental spectral functions
for $t\le s_0$, and the approximation
\begin{equation}
\label{approx}
\r_V(t)-\r_A(t)\approx\r^{\rm DV}_V(t)-\r^{\rm DV}_A(t)\ ,\qquad t\ge s_0\ ,
\end{equation}
above $s_0$, \seef\ Eq.~(\ref{WSR2}).   This approximation involves the assumption that 
OPE contributions in principle present in the theoretical
representation of $\r_V-\r_A$ are numerically tiny and
can be safely neglected.
We can test this assumption by
evaluating the
OPE corrections in Eq.~(\ref{alt}), which in that equation appear as the terms depending on $\a_s(s_0)$.
Setting $s_0=m_\t^2$ and using $\a_s(m_\t^2)/\p\approx 0.1$ and
$m_{u,d}(m_\t^2)<10$~MeV, we find that the second term inside the
square brackets is at most of order $10^{-5}$.  The 
term proportional to $m_\p^2\a_s(s_0)/s_0$ inside the square brackets is of order
$4\times 10^{-4}$ at the $\t$ mass.   For values of $s_0$ down to 1.5~GeV$^2$ it will be larger, but even an order of magnitude will not affect our
results below.\footnote{We will therefore also not worry about higher-order
corrections in $\a_s$ omitted from Eqs.~(\ref{WSR2}) and~(\ref{alt}) above, even though typically the perturbative expansions
of the coefficients $C_{2k,V-A}$ converge slowly for the $J=0$ component.}
In fact, the contribution from the term $2f_\p^2 m_\p^2/s_0^2$ to
$L_{10}^{\rm eff}$ itself
is very small, about $2\times 10^{-5}$.
For our purposes,
the dimension two and four OPE corrections to the approximation~(\ref{approx}) turn out
to be completely negligible, and it will be justified to drop the terms
in Eq.~(\ref{alt})
containing factors of $\a_s(s_0)$ in Sec.~\ref{results} below.\footnote{A less
quantitative version of this argument appeared in Ref.~\cite{GPP2}.}

For the DV part of the vector and axial spectral functions, we will
use the parametrization 
\begin{equation}
\label{DVpar}
\r^{\rm DV}_{V/A}(t)=e^{-\d_{V/A}-\g_{V/A}t}\sin{\left(\a_{V/A}+\b_{V/A}t\right)}\ ,
\end{equation}
where $\a_{V/A}$, $\b_{V/A}$, $\g_{V/A}$, and $\d_{V/A}$ are eight free
DV parameters, which are fitted to moments of the experimental spectral functions.   For a detailed discussion and history of this
parametrization, see Refs.~\cite{CGP05,CGP08,russians}.

\section{\label{data} Strategy and data}
We will evaluate $\P_{V-A}(Q^2)$
and $\P^{(w_k)}_{V-A}(Q^2)$ using OPAL experimental data \cite{OPAL} for the spectral functions $\r_V(t)$ and $\r_A(t)$ for $t\le s_0=t_{\rm switch}$, and
approximating the difference $\r_V(t)-\r_A(t)$ by Eq.~(\ref{approx}) for
$t\ge s_0=t_{\rm switch}$, with values for the DV parameters from our previous fits to the data.   We used 
adjusted OPAL
data, updated to reflect current values of exclusive
mode hadronic $\tau$-decay branching fractions, as described in Ref.~\cite{US2}.  We will choose
$t_{\rm switch}$ to be the upper end of OPAL bin $N$, obtaining
\begin{eqnarray}
\label{PiOPAL}
\P_{V-A}^{(w)}(Q^2)&=&\sum_{i=1}^N\D t \;w(t[i]/t_{\rm switch})\;
\frac{\r_V(t[i])-\r_A(t[i])}{t[i]+Q^2}\\
&&+\int_{t_{\rm switch}}^\infty dt\;w(t/t_{\rm switch})\;\frac{\r^{\rm DV}_V(t)-\r^{\rm DV}_A(t)}{t+Q^2}\ .\nonumber
\end{eqnarray}
Here $\D t=0.032$~GeV$^2$ is the OPAL bin width and $t[i]=(i-1/2)\D t$ is the midpoint
value of the $i$th bin; $t_{\rm switch}=t[N]+\D t/2=N\D t$.  $\P_{V-A}(Q^2)$ is obtained by setting the polynomial weight $w=1$.

The simplest fits from which the DV parameters were obtained were fits
to the separate vector and axial versions of Eq.~(\ref{cauchy}) with $w(t)=1$,
using OPAL data to evaluate the moments
\begin{equation}
\label{moment}
I_{V/A}(s_0)=\int_0^{s_0}dt\;\r_{V/A}(t)
\end{equation}
through a Riemann-sum approximation like the one shown in Eq.~(\ref{PiOPAL}), and
varying $s_0$ between a given $s_{min}$ and $m_\t^2$.  
For $w(t)=1$, all OPE contributions
except the $D=0$ perturbative ones are negligible, and a fit to
$I_{V/A}(s_0)$ thus yields $\a_s$ and the DV parameters of the
channel in question.\footnote{$\a_s$ was enforced to be equal in the two
channels.}  The value of $s_{min}$ was determined by
requiring a good quality match between the experimental $I_{V/A}(s_0)$
and fitted theoretical representations, and stability of the fit parameters
with respect to variation of $s_{min}$.
In this
article, we will always choose $t_{\rm switch}=s_{min}$.\footnote{We have
explored taking $t_{\rm switch}>s_{min}$, and find that this leads to
results fully consistent with the choice $t_{\rm switch}=s_{min}$ and no
reduction in errors.}
Our
central results were obtained with the choice $s_{min}=1.504$~GeV$^2$.\footnote{This value corresponds to the upper end of OPAL bin 47.}
We have also used the more elaborate moments with weights $1-(t/s_0)^2$
and the ``$\t$ kinematic weight'' $(1-t/s_0)^2(1+2t/s_0)$ inserted into
Eq.~(\ref{moment}); the perturbative part of all moments was evaluated
using both fixed-order (FOPT) and contour-improved (CIPT) \cite{CIPT}
perturbation theory.   The non-trivially weighted moments also give access to the
OPE coefficients $C_{6,V/A}$ and $C_{8,V/A}$.   Both pure vector and combined vector and axial channel fits were investigated.
For a detailed account
of all these fits, we refer to Refs.~\cite{US2,US1}.   The fit results employed here
are always those from Ref.~\cite{US2}, unless otherwise noted.

We have fully propagated all errors and correlations in the results we will report on below.
In particular, the DV parameter values used in Eq.~(\ref{PiOPAL}) are
correlated with the data, and we have computed these correlations using the
linear error propagation method summarized in the appendix of Ref.~\cite{US1}
(see, in particular, Eq.~(A.4) of that reference, which can be used to express
the parameter-data covariances in terms of the data covariance matrix).

\section{\label{results} Results}
We will begin with presenting the results for $L^{\rm eff}_{10}$ and $C^{\rm eff}_{87}$ as defined by Eq.~(\ref{lowQ2}), using Eq.~(\ref{alt}) as well.   After that,
we will check the convergence of chiral perturbation theory by
fitting the $Q^2$ dependence to the two-loop expressions for $\P_{V-A}$  calculated in Ref.~\cite{ABT}.   Then, in Sec.~\ref{OPE}, we will revisit
the dimension 6 and 8 OPE coefficients.

\begin{table}[t]
\begin{center}
\vspace*{2ex}
\begin{tabular}{|c|c|c|c|c|c|c|c|}
\hline
$s_{min}$ & $\Pi_{V-A}(0)$ & $\Pi^{\rm DV}_{V-A}(0)$ & $\Pi^{(w_1)}_{V-A}(0)$ & $\Pi^{(w_1){\rm DV}}_{V-A}(0)$ & $\Pi^{(w_2)}_{V-A}(0)$  &  $\Pi^{(w_2){\rm DV}}_{V-A}(0)$   \\
\hline
1.408 & 0.0522(10) & $-0.0039$ & 0.04019(88) & $-0.00042$ & 0.02738(72) & 0.00028 \\
1.504 & 0.0522(11) & $-0.0019$ & 0.04083(90) & $-0.00071$ & 0.02915(72) & 0.00033 \\
1.600 & 0.0523(11) & $-0.0001$ & 0.04081(90) & $-0.00066$ & 0.02916(72) & 0.00013 \\
\hline
\hline
1.504 & 0.0522(11) & $-0.0019$ & 0.04081(91) & $-0.00072$ & 0.02916(72) & 0.00034 \\
\hline
\end{tabular}
\end{center}
\vspace*{4ex}
\caption{Values of $\P_{V-A}$, $\P^{(w_1)}_{V-A}$, and $\P^{(w_2)}_{V-A}$ at $Q^2=0$.  We always take the
switch point between data and the duality-violating part of the spectral
function at $t_{\rm switch}=s_{min}$ (values for $s_{min}$ are in GeV$^2$).  The superscript DV indicates the contribution from the second term on the right-hand side of Eq.~(\ref{PiOPAL}).   Duality violation parameters are from the
fits of Ref.~\cite{US2}, Table~3.  Results from fits using FOPT are shown above the double line, those from CIPT below.}
\label{t1}
\end{table}%
\begin{boldmath}
\subsection{\label{LECs} $L^{\rm eff}_{10}$ and $C^{\rm eff}_{87}$}
\end{boldmath}
Table~\ref{t1} shows results relevant for $L^{\rm eff}_{10}$.  This LEC
can be directly obtained from the second column using Eq.~(\ref{lowQ2}),
or from the fourth or sixth column using Eq.~(\ref{alt}).   The DV parts of
these integrals, corresponding to the second term on the right-hand side of Eq.~(\ref{PiOPAL}),
are shown in the third, fifth and seventh columns.   Note that the (absolute)
errors become smaller with increasing $k$ in Eq.~(\ref{weights}), \ie, with
more pinching at $s_{min}=t_{\rm switch}$.   We also note that the results
are essentially independent of $s_{min}$, and whether one chooses
the FOPT or CIPT scheme for the evaluation of the truncated perturbative
series. This is a consequence of the fact that the integrals are almost completely
determined by the data part, \ie, the sum on the first line of Eq.~(\ref{PiOPAL}),
as can be seen from the always small contribution from the DV part of the integrals.
We will henceforth use the FOPT results at $s_{min}=1.504$~GeV$^2$.   

{}From Eq.~(\ref{lowQ2}) we find
\begin{equation}
\label{L10a}
L^{\rm eff}_{10}=(-6.52\pm 0.14)\times 10^{-3}\qquad  (\mbox{from}\ \Pi_{V-A}(0))\ .
\end{equation}
Using Eq.~(\ref{alt}), one may also compute $L^{\rm eff}_{10}$ from the
other values shown in Table~\ref{t1}; the results are always consistent
within errors.   In fact, using $\P^{(w_{1,2})}_{V-A}$ and Eq.~(\ref{alt}), we obtain 
the somewhat more precise values:
\begin{subequations}
\label{L10b}
\begin{eqnarray}
L^{\rm eff}_{10}&=&(-6.52\pm 0.11)\times 10^{-3}\qquad  (\mbox{from}\ \Pi^{(w_1)}_{V-A}(0))\ ,\label{L10ba}\\
&=&(-6.45\pm 0.09)\times 10^{-3}\qquad  (\mbox{from}\ \Pi^{(w_2)}_{V-A}(0))\ .\label{L10bb}
\end{eqnarray}
\end{subequations}

These values are in good agreement with the value found recently in
Ref.~\cite{GPP3}, except that our best error is twice as large.   There are 
(at least) two
reasons for this difference in errors, both of which point to the error in
Ref.~\cite{GPP3} being underestimated.\footnote{For more comments on the
comparison with Ref.~\cite{GPP3}, we refer to the Conclusion.}    First, Refs.~\cite{GPP2,GPP3} used
a DV {\it ansatz} of the functional form shown in
Eq.~(\ref{DVpar}) for the {\it difference} $\r^{\rm DV}_V-\r^{\rm DV}_A$, instead
of using this form for each channel separately.   That implies that Refs.~\cite{GPP2,GPP3} used only four parameters to describe
duality violations in $V-A$, whereas we use eight.   
The simplified four-parameter form
assumed in Refs.~\cite{GPP2,GPP3} would be valid if it happened,
for some reason, that $\g_V=\g_A$ and $\b_V=\b_A$.
Since we find very
different values for $\g_V$ and $\g_A$ in our fits to both the OPAL data
\cite{US2} and the ALEPH data \cite{CGP08f}, this condition
is, however, not satisfied. 
The theoretical systematic error associated
with the breakdown of this assumption 
is, of course, not
included in the error estimates of Refs.~\cite{GPP2,GPP3}.
These comments
remain relevant  even if an {\it ansatz} of the form~(\ref{DVpar}) gives a reasonable description
of the difference $\r_V(t)-\r_A(t)$ for large enough $t$:  a model description
of duality violations is only acceptable if it describes the resonance physics
at higher energies in both the vector and axial channels individually.
The second reason for our larger error is that Ref.~\cite{GPP3}
used the formally more precise, but in practice incomplete ALEPH data \cite{TAU10}.
If ALEPH data with corrected correlation matrices were to become available, we anticipate that errors
would be reduced relative to those obtained using the OPAL data
for our fits as well.

\begin{table}[t]
\begin{center}
\vspace*{2ex}
\begin{tabular}{|c|c|c|}
\hline
$s_{min}$  &   $\Pi'_{V-A}(0)$ & ${\Pi'}^{\rm DV}_{V-A}(0)$    \\
\hline
1.408 &  $-0.1356(47)$ &  0.0029  \\
1.504 &  $-0.1355(47)$ &  0.0016  \\
1.600 &  $-0.1356(47)$ &  0.0004  \\
\hline
\hline
1.504 &  $-0.1355(47)$ &  0.0016  \\
\hline
\end{tabular}
\end{center}
\vspace*{4ex}
\caption{Values of $\P'_{V-A}$ at $Q^2=0$
obtained by differentiating
Eq.~(\ref{defPiLR}) with respect to $Q^2$.  We always take the
switch point between data and the duality-violating part of the spectral
function at $t_{\rm switch}=s_{min}$.  The superscript DV indicates the contribution from the second term on the right-hand side of Eq.~(\ref{PiOPAL}).
Duality violation parameters are from the
fits of Ref.~\cite{US2}, Table~3.  Results from fits using FOPT are shown above the double line, those from CIPT below.}
\label{t2}
\end{table}%

Values for the derivative of $\P_{V-A}(Q^2)$ with respect to $Q^2$ at
$Q^2=0$ are shown in Table~\ref{t2}.   As one would expect, the results
show the same robustness with respect to the various fits of Ref.~\cite{US2} 
as those in Table~\ref{t1}.   Using Eq.~(\ref{PiOPAL}), we find
\begin{equation}
\label{C87}
C_{87}^{\rm eff}=(8.47\pm 0.29)\times 10^{-3}\ \mbox{GeV}^{-2}\ .
\end{equation}
This value again agrees with that found in Ref.~\cite{GPP3}, but our error is
again about twice as large.   Using the cubic doubly-pinched weight of Ref.~\cite{MY} in Eq.~(\ref{defPiLRw})
as was done in Ref.~\cite{GPP3} does not lead to a smaller error in our case.
The same comments about the reasons for our larger error as discussed
above for $L_{10}^{\rm eff}$ apply here as well.

We have repeated the analysis presented here using fit values for the
DV parameters reported in Table~5 of Ref.~\cite{US2}, again taking all 
correlations into account.  The results for $L_{10}^{\rm eff}$ and
$C_{87}^{\rm eff}$ are virtually identical to those reported above.

\subsection{\label{ChPT} Connection to chiral perturbation theory}
The LECs $L_{10}^{\rm eff}$ and $C_{87}^{\rm eff}$, which are 
defined by the values at $Q^2=0$ of $\P_{V-A}(Q^2)$ and its
derivative (\seef\ Eq.~(\ref{lowQ2})), are connected to 
LECs in the order-$p^6$ chiral lagrangian through the relations \cite{GPP}
\begin{subequations}
\label{chptconn}
\begin{eqnarray}
\P_{V-A}(0)&=&-8L_{10}^{\rm eff}\label{chptconna}\\
&=&-8L_{10}^r(\m)\Bigl(1-4(2\m_\p+\m_K)\Bigr)
+16(2\m_\p+\m_K)L_9^r(\m)\nonumber\\
&&-\frac{1}{16\p^2}\left(1-\log\frac{\m^2}{m_\p^2}+\frac{1}{3}\log\frac{m_K^2}{m_\p^2}\right)-8G_{2L}(\m,0)\nonumber\\
&&-32m_\p^2\left(C_{61}^r(\m)-C_{12}^r(\m)-C_{80}^r(\m)\right)\nonumber\\
&&-32(2m_K^2+m_\p^2)\left(C_{62}^r(\m)-C_{13}^r(\m)-C_{81}^r(\m)\right)
\ ,\nonumber
\\
-\P_{V-A}'(0)&=&16C_{87}^{\rm eff}\label{chptconnb}\\
&=&16C_{87}^r(\m)+\frac{1}{480\p^2}\left(\frac{1}{m_\p^2}+\frac{2}{m_K^2}\right)-8\,\frac{\partial G_{2L}(\m,s)}{\partial s}\Bigg|_{s=0}\nonumber\\
&&-\frac{1}{4\p^2f_\p^2}\left(1-\log\frac{\m^2}{m_\p^2}+\frac{1}{3}\log\frac{m_K^2}{m_\p^2}\right)L_9^r(\m)
\ ,\nonumber\\
\m_P&=&\frac{m_P^2}{32\p^2f_\p^2}\log\frac{m_P^2}{\m^2}\ .\label{chptconnc}
\end{eqnarray}
\end{subequations}
Here the superscript $r$ denotes the values of LECs renormalized at scale
$\m$, which below we will take to be $\m=0.77$~GeV.

The complete order-$p^6$ ChPT expression for $\P_{V-A}(Q^2)$ can be written as a function of $Q^2$ in terms of the renormalized LECs $L_{9,10}^r$ and $C_{12,13,61,62,80,81,87}^r$ using the results of Ref.~\cite{ABT}.\footnote{We do not quote those results here because of their length.} 
Choosing $\m=0.77$~GeV, and using $m_\p=139.570$~MeV and $m_K = 495.65$~MeV, the $Q^2$ dependence of $\P_{V-A}(Q^2)$ in 
chiral perturbation theory to order $p^6$ takes the form
\begin{eqnarray}
\label{Pifit}
\P_{V-A}(Q^2)&=&-12.165L_{10}^r-32m_\p^2\left(C_{61}^r-C_{12}^r-C_{80}^r\right)\\
&&-32(2m_K^2+m_\p^2)\left(C_{62}^r-C_{13}^r-C_{81}^r\right)-16C_{87}^rQ^2+R(Q^2;L_9^r)\ ,\nonumber
\end{eqnarray}
where $R(Q^2;L_9^r)$ is a fully known non-analytic function in $Q^2$ coming from one- and
two-loop contributions in ChPT, including one-loop contributions with a
vertex containing $L_9^r$.   Note that $R(Q^2;L_9^r)$ also depends on the
scale $\m$, even though we have not explicitly indicated any such dependence in Eq.~(\ref{Pifit}), because we evaluated the numerical value
of the coefficient of $L_{10}^r$ at  $\m=0.77$~GeV.   This implies that
both $R(Q^2;L_9^r)$ and all LECs appearing in this equation are to be
evaluated at this value of $\m$.
At $Q^2=0$, Eq.~(\ref{Pifit})
yields Eq.~(\ref{chptconn}), through the relation~(\ref{lowQ2}).

If we fit $\P_{V-A}(Q^2)$ to this order-$p^6$ expression, we can explore the
range in $Q^2$ for which order-$p^6$ ChPT is a valid approximation.   Note
that the order-$p^6$ expression is not linear in $Q^2$, even though Eq.~(\ref{lowQ2}), which one obtains upon
re-expanding the order-$p^6$ ChPT expression for $Q^2<4m_\p^2$, is linear
in $Q^2$.\footnote{The threshold in the dispersive integral for $\P_{V-A}(Q^2)$ is $4m_\p^2$, and not $m_\p^2$, since the $\p$ pole contribution was subtracted in defining
$\r_A(t)$.}
With 
$m_\p$ and $m_K$ fixed to their physical values, the data can, of course, 
not be used to separate the
$Q^2$-independent part of Eq.~(\ref{Pifit}) into its individual order-$p^4$ and order-$p^6$ components without
additional input.  Such input can, in principle, be obtained from lattice
studies employing a range of light quark masses.

We have carried out a fit to Eq.~(\ref{Pifit}) in terms of $L_{10}^r$ and $C_{87}^r$, using
given values for all the other LECs on the right-hand side of Eq.~(\ref{chptconn}).
Specifically, we chose the central values $L_9^r(\m)=0.00593$
\cite{BT}, $4m_\p^2(C_{61}^r(\m)-C_{12}^r(\m)-C_{80}^r(\m))=-0.000067$
and $C_{62}^r(\m)-C_{13}^r(\m)-C_{81}^r(\m)=0$ \cite{GPP} at $\m=0.77$~GeV,\footnote{Errors on these values are only needed if one wishes
to convert values for $L_{10}^{\rm eff}$ and $C_{87}^{\rm eff}$ into values
for $L_{10}^r$ and $C_{87}^r$.  For such an analysis, we refer to Sec.~\ref{LECvalues}.} for which Eq.~(\ref{chptconn}) becomes
\begin{subequations}
\label{chptconnnum}
\begin{eqnarray}
L_{10}^{\rm eff}&=&1.521L_{10}^r(\m=0.77\ {\rm GeV})-0.000288\ ,
\label{chptconnnuma}\\
C_{87}^{\rm eff}&=&C_{87}^r(\m=0.77\ {\rm GeV})+0.00328\ \mbox{GeV}^{-2}\ ,
\label{chptconnnumb}
\end{eqnarray}
\end{subequations}
where we also used
$m_\eta=547.853$~MeV (the latter is needed for the evaluation of the loop
contributions to the constants in these equations).

\begin{figure}[t]
\centering
\includegraphics[scale=1.1]{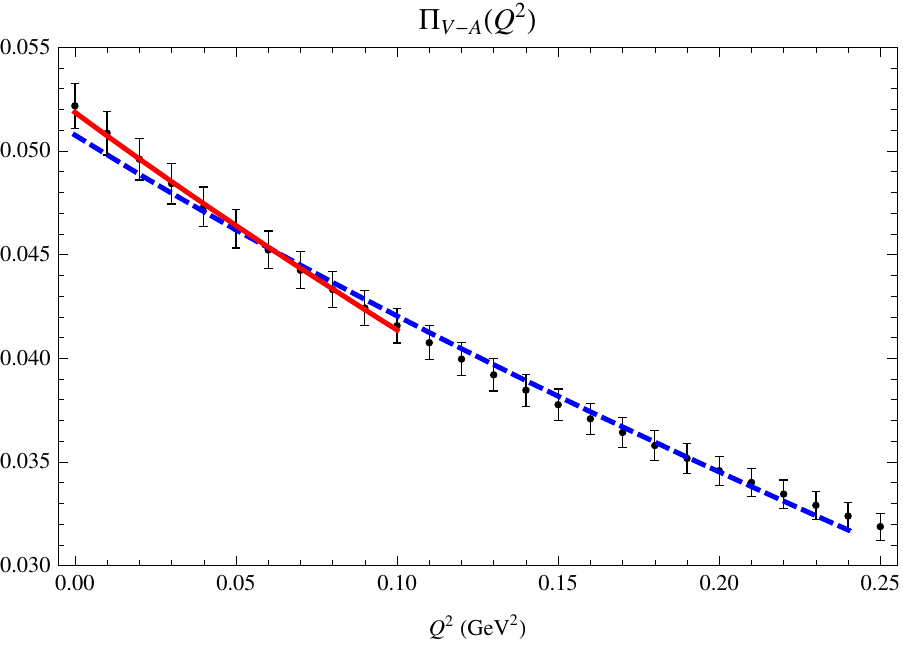}
\floatcaption{f1}{ChPT fits at order $p^6$ to $\P_{V-A}(Q^2)$.   The blue
(dashed)
curve includes $Q^2$ values up to 0.24~GeV$^2$; the red (continuous) curve
includes $Q^2$ values up to 0.10~GeV$^2$.}
\vspace*{2ex}
\end{figure}

Fits to ChPT at order $p^6$ are shown in Fig.~\ref{f1}.   The blue (dashed) curve
shows a fit with a maximum $Q^2$ value  $Q^2_{max}=0.24$~GeV$^2$, while the red
(continuous)
curve shows a fit with  $Q^2_{max}=0.10$~GeV$^2$.   
The fits were performed using points beginning at $Q^2=0$ and spaced by 0.01~GeV$^2$.
These data, computed from Eq.~(\ref{PiOPAL}), are strongly correlated,
and not amenable to a standard $\c^2$ fit, forcing us to perform
a fit with diagonal inverse-squared-error weighting.\footnote{Thinning out the
data does not help.} The full
data correlations are then taken into account in the quoted
errors using the technique described in the appendix of Ref.~\cite{US1}.

Clearly, the blue curve does not provide a good fit, while the red curve does.
We conclude that ChPT at this order gives a good match to $\P_{V-A}(Q^2)$
up to $Q\approx 300$~MeV, which is about twice the pion mass.   The ChPT
fits are virtually linear, suggesting consistency with the  extraction of 
$L_{10}^{\rm eff}$ and $C_{87}^{\rm eff}$ from Eq.~(\ref{lowQ2}).   From the
ChPT fit corresponding to the red curve in Fig.~\ref{f1}, we obtain the values
$L_{10}^r=-4.08(9)\times 10^{-3}$ and $C_{87}^r=3.97(18)\times 10^{-3}$~GeV$^{-2}$, which correspond to
\begin{eqnarray}
\label{chptvalues}
L_{10}^{\rm eff}&=&(-6.49\pm 0.14)\times 10^{-3}\ ,\\
C_{87}^{\rm eff}&=&(7.25\pm 0.18)\times 10^{-3}\ \mbox{GeV}^{-2}\ .\nonumber
\end{eqnarray}
The value for $L_{10}^{\rm eff}$ is completely consistent with Eqs.~(\ref{L10a})
and~(\ref{L10b}), but this is not the case for the value of $C_{87}^{\rm eff}$,
which is not consistent
within errors with Eq.~(\ref{C87}).   The reason for this is that the value in
Eq.~(\ref{C87}) was obtained from the behavior of $\P_{V-A}(Q^2)$ near
$Q^2=0$, while the value in Eq.~(\ref{chptvalues}) was obtained by a fit
of $\P_{V-A}(Q^2)$ over the range $0\le Q^2\le Q^2_{max}=0.1$~GeV$^2$.
Values for $C_{87}^{\rm eff}$ obtained by varying $Q^2_{max}$
are shown as the black points (crosses) in Fig.~\ref{f2}.   This figure shows that ChPT to order $p^6$
does a reasonable job in describing $\P_{V-A}(Q^2)$, but clearly
order-$p^8$ effects, not included in the chiral fits, are present in the data.
In contrast, the value of $L_{10}^{\rm eff}$ is barely affected by varying
$Q^2_{max}$; it varies by less than the errors quoted in Eq.~(\ref{L10b})
over the range shown in Fig.~\ref{f2}.

\begin{figure}[t]
\centering
\includegraphics[scale=1.1]{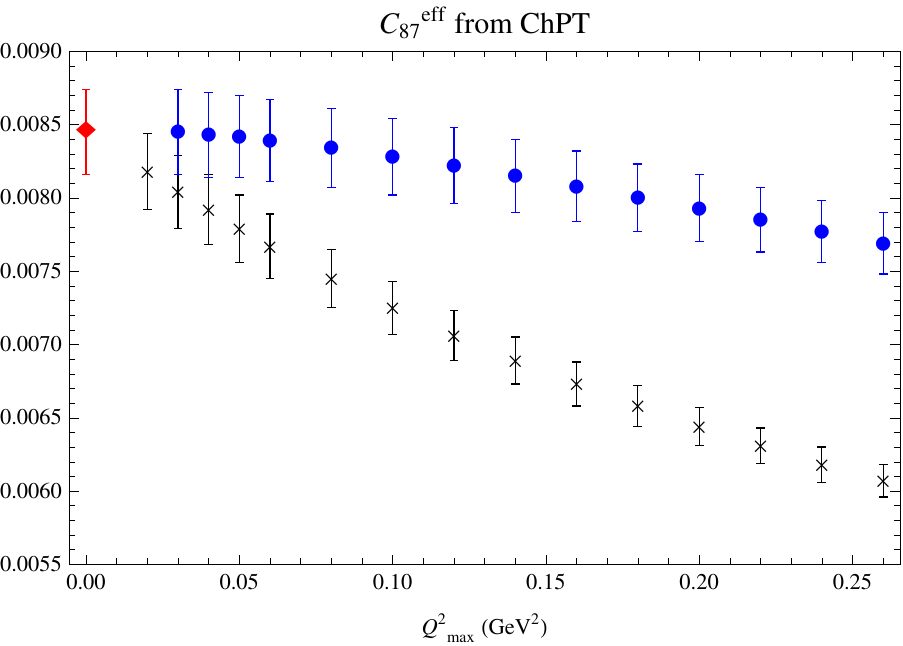}
\floatcaption{f2}{Black points (crosses) show values of $C_{87}^{\rm eff}$ obtained from the ChPT fits
as a function of $Q^2_{max}$, the maximum $Q^2$ value used in the fit.
The red point (diamond) at $Q^2_{max}=0$ is the value of Eq.~(\ref{C87}), for comparison.
The blue points (filled circles) have been obtained from a fit to Eq.~(\ref{Pifit}) with a
term proportional to $Q^4$ added to it; see text for further details.}
\vspace*{2ex}
\end{figure}

The presence of order-$p^8$ effects can be checked by redoing the ChPT fits, but now using Eq.~(\ref{Pifit})
with an extra term $+DQ^4$ added.   This is of course a phenomenological fit,
because the order-$p^8$ structure is more complicated than just such a 
simple term.  But Fig.~\ref{f1} shows that the deteriorating quality of the
fits with larger values of $Q^2_{max}$ is due to some curvature showing
up in $\P_{V-A}(Q^2)$ at larger $Q^2$, and we expect this extra term to
capture this curvature reasonably well.   We show the results for
$C_{87}^{\rm eff}$ as a function of $Q^2_{max}$ with this new term included
in the fit
as the blue points (filled
circles) in Fig.~\ref{f2}.  Indeed, the values for $C_{87}^{\rm eff}$ become much less
sensitive to $Q^2_{max}$, with values consistent with Eq.~(\ref{C87}) over
a much larger range.   We also find that $L_{10}^{\rm eff}$
does not change significantly as a consequence of this exercise:
instead of the value in Eq.~(\ref{chptvalues}) we now obtain
$L_{10}^{\rm eff}=(-6.52\pm 0.14)\times 10^{-3}$.  
The phenomenological coefficient $D$ varies between 0.2 and 0.1 over
the interval shown in the figure.

The lesson of this exploration is that any values of $L_{10}^r$ and
$C_{87}^r$ obtained from $L_{10}^{\rm eff}$ and $C_{87}^{\rm eff}$ using (as
in Ref.~\cite{GPP}) the order-$p^6$ ChPT relations of Eq.~(\ref{chptconn}) must be treated with
some care. While terms beyond order-$p^6$ in the chiral counting associated
with higher powers of $Q^2$ can be removed by taking $Q^2$ to zero,
those associated with higher powers of the quark masses are fixed by
the non-zero, physical meson masses and cannot be removed. Such
contributions are present in the relations between $L_{10}^{\rm eff}$ and
$L_{10}^r$ and $C_{87}^{\rm eff}$ and $C_{87}^r$ to arbitrarily
high chiral order. The variation in the fitted value of $C_{87}^r$
with $Q^2_{max}$ (the source of the variation of $C_{87}^{\rm eff}$
displayed in Fig.~\ref{f2}, \seef\ Eq.~(\ref{chptconnnumb})) indicates non-trivial $Q^2$-dependent contributions
of order-$p^8$ and beyond, raising the possibility of analogous 
mass-dependent, $Q^2$-independent order-$p^8$ (and beyond) contributions
as well. 

The impact of such order-$p^8$ (and higher) contributions will be
more significant for the relation between $C_{87}^{\rm eff}$ and $C_{87}^r$
than for that between $L_{10}^{\rm eff}$ and $L_{10}^r$ since in the
former case the missing order-$p^8$ terms are only one chiral
order higher than the LEC of interest, $C_{87}^r$, whereas in the
latter case the missing terms begin two chiral orders higher than
$L_{10}^r$. Even so, the order-$p^8$ and higher contributions need not
be completely negligible for $L_{10}^r$. In fact, ignoring
the contributions of the order-$p^6$ LECs $C_{12,13,61,62,80,81}$
to the relation between $L_{10}^{\rm eff}$ and $L_{10}^r$, the effects of
the mass-dependent order-$p^6$ terms are significant, changing the coefficient
of $L_{10}^r$ from $1$ at order-$p^4$ to $1.521$ at order-$p^6$
in Eq.~(\ref{chptconnnuma}), and altering the
best fit results for $L_{10}^r$ by about $30\%$ between order-$p^4$ and order-$p^6$.
A further shift in $L_{10}^r$ by about $0.3^2\approx 10\%$ due to order-$p^8$
effects would thus not be unexpected. Similarly, a difference of about
$30\%$ between the order-$p^6$ and order-$p^8$ values for $C_{87}^r$ would
not be surprising. 

In conclusion, if estimates for $L_{10}^r$ and $C_{87}^r$ obtained
from $L_{10}^{\rm eff}$ and $C_{87}^{\rm eff}$
are used in the computation of some other physical
quantity, propagating the error quoted in Eqs.~(\ref{L10b}) and~(\ref{C87}) would not include
additional systematic errors due to the omission of order-$p^8$ terms
in ChPT.    This general perspective applies, of course, to the next
subsection, in which we attempt to extract values for $L_{10}^r$ and
$C_{87}^r$ from our analysis.

\begin{boldmath}
\subsection{\label{LECvalues} Estimates for $L^r_{10}$ and $C^r_{87}$}
\end{boldmath}
In this subsection, we convert the values of Eqs.~(\ref{L10bb}) and~(\ref{C87}) for 
$L^{\rm eff}_{10}$ and $C^{\rm eff}_{87}$ into values for $L^r_{10}(\m)$ and
$C^r_{87}(\m)$.  We will leave the $\m$ dependence of
all LECs implicit, where it is to be understood that all numerical values have
been evaluated at $\m=0.77$~GeV.
{}From Eq.~(\ref{chptconna}), it is clear that this requires input on the two order-$p^6$ LEC combinations,
\begin{eqnarray}
C_0 &\equiv&
32 m_\pi^2 \left( C^r_{12}-C^r_{61}+C^r_{80}\right)\ ,\nonumber\\
C_1&\equiv& 32 \left( m_\pi^2+2m_K^2\right)
\left( C^r_{13}-C^r_{62}+C^r_{81}\right)\ .
\label{c0c1defns}
\end{eqnarray}
The results for $\Pi_{V-A}(0)$ we obtained above from the $\t$ spectral functions
correspond to the rather strongly constrained relation,
\begin{equation}
L_{10}^r= -0.004143(89)_{OPAL}(74)_{L_9^r}
\, +\, 0.0822(C_0+C_1)\ ,
\label{deltapibarzeroconstraint}
\end{equation}
where the first component of the error on the right-hand side is experimental
and the second that due to the uncertainty on the input employed
for $L_9^r$~\cite{BT}.

While the LECs in $C_0$ are all zeroth order in $1/N_c$
and those in $C_1$ first order, the ratio
$(m_\pi^2+2m_K^2)/m_\pi^2\simeq 26$ of factors multiplying the LECs in
$C_1$ and $C_0$ more than compensates for the $1/N_c$
suppression, potentially making $C_1$ the numerically more important of the two.
Unfortunately, while some estimates exist for the LECs entering $C_0$, nothing is
known of those entering $C_1$.

In Ref.~\cite{GPP}, this situation was handled as follows.  The combination
$C_0$ was first determined using existing estimates
of $C_{12}^r$~\cite{jop04c12}, $C_{61}^r$~\cite{dk00c61,km06c61}
and $C_{80}^r$~\cite{up08c80}. The combination $C_1$, for which no 
analogous
estimates exist,
was then set to zero and assigned an error based on the assumption
\begin{equation}
\vert C_{13}^r-C_{62}^r+C_{81}^r\vert < {\frac{1}{3}}\,
\vert  C^r_{12}-C^r_{61}+C^r_{80}\vert\ ,
\label{gappc1}
\end{equation}
the $1/3$ on the RHS reflecting the $1/N_c$ suppression. The
uncertainty on $L_{10}^r$ reported in Ref.~\cite{GPP} is
entirely dominated by the resulting error on $C_1$.
It is thus relevant to assess whether or not
this assumption is a sufficiently conservative one.

We consider first the input values employed on the right-hand side of
Eq.~(\ref{gappc1}).  The value $C_{12}^r=(0.4\pm 6.3)\times 10^{-5}$~GeV$^{-2}$ has
been determined from a highly constrained, mildly model-dependent
treatment of the $K\pi$ scalar form factor \cite{jop04c12}.\footnote{Note that the definitions of $C_{12}^r$ here and in
Ref.~\cite{jop04c12} differ by a factor of $f_\p^2$. Furthermore,
even though the  input values employed in Ref.~\cite{jop04c12} for both $f_K/f_\p$
and $F_+(0)$ were somewhat different from modern values, the
corresponding shifts in $C_{12}^r$ largely cancel such that it
practically stays the same.}
This value is in rough agreement with estimates obtained in
the Resonance Chiral Perturbation Theory (RChPT)
model~\cite{bt03c12,vcetal05c12}.
RChPT estimates also exist for $C_{61}^r$~\cite{ABT,km06c61}
and $C_{80}^r$~\cite{ABT,up08c80}. The $C_{61}^r$ and $C_{80}^r$
estimates of Ref.~\cite{ABT} are numerically equal, as are the
$C_{61}^r$ estimate of Ref.~\cite{km06c61} and $C_{80}^r$ estimate of
Ref.~\cite{up08c80}. One thus expects significant cancellation
between the $C_{61}^r$ and $C_{80}^r$ contributions to $C_0$.
To the best of our knowledge, the RChPT estimates of Ref.~\cite{ABT,up08c80}
are the only sources of information on $C_{80}^r$. Averaging the
two central values yields $C_{80}^r=(2.0\pm 0.5)
\times 10^{-3}$~GeV$^{-2}$, the error reflecting only the uncertainties
on experimental inputs to the underlying RChPT fits, and not the systematic
error from the use of RChPT.
Finally, $C_{61}^r$ has been determined from an inverse-weighted
finite-energy sum rule involving the difference of non-strange and
strange vector-channel spectral functions measured in hadronic $\t$
decays~\cite{dk00c61,km06c61}.\footnote{The result quoted in
Ref.~\cite{km06c61} is actually supposed to represent, up to a change
in notation, that obtained in Ref.~\cite{dk00c61}. Owing to a sign
transcription error, however, the result employed for the difference of the
non-strange and strange correlators at $Q^2=0$, needed in the
evaluation of $C_{61}^r$, has been inadvertently shifted, altering
the result for $C_{61}^r$. The original  result of Ref.~\cite{dk00c61}
corresponds to $C_{61}^r=(8.1\pm 3.9)\times 10^{-4}$~GeV$^{-2}$. We thank Bachir Moussallam for clarifying this point.}
Updating the input to that analysis, to reflect current values of various
input parameters which differ significantly from those available
at the time Ref.~\cite{dk00c61} appeared, and using the above values
for $C_{12}^r$ and $C_{80}^r$,\footnote{For $C_{80}^r$, for which, to the best of our knowledge, no
experimental estimate exists, we have used the difference between
the RChPT value and the (updated) experimental value of
$C_{61}^r$ as an estimate of the systematic uncertainty
on $C_{80}^r$ associated with the use of the RChPT framework.
This component has been added in quadrature
to the error obtained in the RChPT fits, already quoted in the text,
to obtain the total error on $C_{80}^r$.}
one finds $C_{61}^r=(1.4\pm 0.3)\times 10^{-3}$~GeV$^{-2}$ \cite{kmprivate}, and
thus
\begin{equation}
C_0=(3.8\pm 5.3)\times 10^{-4}\ .
\label{c0mixed}
\end{equation}

There is, indeed, a rather strong
cancellation between the $C_{61}^r$ and $C_{80}^r$ contributions
to $C_0$. From the RChPT perspective, where the LECs appearing in
$C_0$ receive strong resonance contributions, while those appearing
in $C_1$ do not, there is no reason to suppose that a similar
cancellation will be operative in $C_1$. An alternate,
 more (but still not excessively)
conservative assumption, which avoids presuming any such strong cancellation
in $C_1$, would be
\begin{equation}
\vert C_{13}^r-C_{62}^r+C_{81}^r\vert < {\frac{1}{3}}\,\left[
\vert  C^r_{12}\vert \, +\, \vert C^r_{61}\vert \, +\, \vert C^r_{80}\vert
\right]\ .
\label{gappc1alt}
\end{equation}
This bound, however, is a factor of about 7 larger than that of
Eq.~(\ref{gappc1}), and would lead to a rather large 
uncertainty, $\sim 0.0016$, on $L_{10}^r$, still without any clear sense
of whether the assumption underlying it is a sufficiently conservative one.

An alternative approach to dealing with this problem has been considered in
Ref.~\cite{kmrbcukqcd13}.  The idea is to consider the $m_\p$ and $m_K$
dependence of 
\begin{equation}
\label{LPiVA}
\D\P(Q^2)\equiv \P^{\rm L}_{V-A}(Q^2)-\P_{V-A}(Q^2)\ ,
\end{equation}
the difference between the $V-A$ correlator $\P^{\rm L}_{V-A}(Q^2)$ evaluated on the lattice, for
unphysical values of the pion and kaon masses, and the same correlator for
the physical mass case,  obtained from the $\t$ spectral
functions.   
Since the same combinations
of order-$p^6$ LECs enter the physical and unphysical mass cases,
the difference of the correlators for the two cases
can be written in the form
\begin{equation}
\D\P (Q^2) = \D R^{\rm L}(Q^2)
\, +\, \d^{\rm L}_{10}\, {\rm L}_{10}^r\, +\, \d^{\rm L}_0 C_0\, +\, \d^{\rm L}_1 C_1\ ,
\label{deltadeltapibar}
\end{equation}
where $\D R^{\rm L}(Q^2)$ and the $Q^2$-independent coefficients $\d^{\rm L}_{10,0,1}$ are known in terms of the lattice and physical meson masses and the renormalization scale $\m$.   Of course,  all LECs are mass-independent
(this is also true for the effective order-$p^8$ coefficient $D$, at least to order $p^8$).
Using Eqs.~(\ref{deltapibarzeroconstraint}) and~(\ref{deltadeltapibar}) yields a constraint on $C_0$ and $C_1$
for each set of lattice values for $m_\p$ and $m_K$, as well as each value of $Q^2$, with different $Q^2$ values at constant lattice meson masses
providing self-consistency checks.   This assumes that lattice results have
been extrapolated to the continuum limit; as we will rely
        on preliminary results from Ref.~\cite{kmrbcukqcd13}, which has yet to study
        this issue, we will neglect the effect of non-zero
        lattice spacing.

In Ref.~\cite{kmrbcukqcd13}, these constraints have been analyzed
for RBC/UKQCD $n_f=2+1$ DWF ensembles with $a^{-1}=1.37$~GeV,
and $m_\p= 171, \, 248$~MeV~\cite{rbcukqcdcoarse12}
and $a^{-1}=2.28$~GeV and $m_\p = 289,\, 344$~MeV~\cite{rbcukqcdfine11},
leading to the preliminary result\footnote{In Ref.~\cite{kmrbcukqcd13} only the pion mass varies significantly, with
the kaon mass staying within 15\% of its physical value \cite{rbcukqcdcoarse12,rbcukqcdfine11}.
We thank the authors of Ref.~\cite{kmrbcukqcd13} for making their preliminary results on $C_0+C_1$
available to us in advance of publication.}
\begin{equation}
C_0+C_1\, =\, (1.3\pm 1.0)\times 10^{-2}\ .
\label{c0c1lattice}
\end{equation}
Note that the associated result for $C_0$, $C_0=-(8.1\pm 8.2)\times 10^{-4}$,
agrees with the estimate of Eq.~(\ref{c0mixed}) within
errors, confirming the utility of RChPT in estimating the order of magnitude
for $C_{80}^r$.   Note also that the
central value for $C_1$ is about two times larger
than allowed by the bound~(\ref{gappc1}).\footnote{It is in the range of the more conservative bound~(\ref{gappc1alt}).}  This, of course, is important
for the determination of $L^r_{10}$.

With the lattice result~(\ref{c0c1lattice}) as input, we finally
obtain
\begin{equation}
L_{10}^r(\m=0.77\ \mbox{GeV})\, =\, (-3.1\pm 0.8)\times 10^{-3}\ ,
\label{finall10}
\end{equation}
with the error entirely dominated by that on $C_0+C_1$.   It should be
kept in mind that not all systematic errors associated with the use of lattice
values for $\P_{V-A}(Q^2)$ have been taken into account.

While, to order $p^6$, the determination of $C_{87}^r$ from $C_{87}^{\rm eff}$
does not suffer from the presence of terms analogous to $C_0$ and
$C_1$, such mass-dependent, but $Q^2$-independent,
contributions would appear in $C_{87}^{\rm eff}$ at order $p^8$. In the case
of $L_{10}^r$, including the order-$p^6$ $C_0+C_1$ contribution
using the lattice estimate leads to a $\sim 25\%$ reduction
compared to the value that would be obtained neglecting them.
We take this $\sim 25\%$ shift as being typical of what one might
expect for contributions to $Q^2=0$ quantities from
missing higher-order mass-dependent terms.  We hence assign
an additional $25\%$ uncertainty to the result we find from Eq.~(\ref{chptvalues})
for $C_{87}^r$, which was obtained from an analysis
including $Q^2$-dependent, but not mass-dependent, order-$p^8$ contributions.   Our final result for $C_{87}^r$ thus becomes
\begin{equation}
C_{87}^r(\m=0.77\ \mbox{GeV}) = (4\pm 1)\times 10^{-3}\ \mbox{GeV}^{-2}\ .
\label{finalc87}
\end{equation}

\begin{boldmath}
\subsection{\label{OPE} $V-A$ condensates}
\end{boldmath}
In this subsection, we consider the values of the OPE coefficients $C_{6,V-A}$ and $C_{8,V-A}$, defined in Eq.~(\ref{highQ2}).   In Ref.~\cite{US2} we presented 
fit results for $C_{6,V/A}$ and  $C_{8,V/A}$ obtained using sum rules involving weights up to degree three,
from which it is straightforward to obtain $C_{6,V-A}$ and $C_{8,V-A}$.
{}From the fits at $s_{min}=1.504$~GeV$^2$, and including 
all correlations, we find the values
\begin{subequations}
\label{C6C8}
\begin{eqnarray}
C_{6,V-A} &=&(-10.5\pm 2.8)\times 10^{-3}\ \mbox{GeV}^6 \qquad   (\mbox{FOPT})\ ,\label{C6C8a}\\
          &=& (-11.3\pm 2.4)\times 10^{-3}\ \mbox{GeV}^6  \qquad  (\mbox{CIPT})\ ,\nonumber\\
C_{8,V-A} &=&  (14\pm 7)\times 10^{-3}\ \mbox{GeV}^8 \qquad   (\mbox{FOPT})\ ,\label{C6C8b}\\
          &=&  (16\pm 6)\times 10^{-3}\ \mbox{GeV}^8  \qquad  (\mbox{CIPT})\ .\nonumber
\end{eqnarray}
\end{subequations}
Changes as a function of varying $s_{min}$ are small compared to the
errors shown in Eq.~(\ref{C6C8}).

It is interesting to compare these values with those we would obtain from
the original OPAL data, to which no correction reflecting modern values for the
$\t$ hadronic branching fractions have been applied.   In this case, we find,
using the fits reported in Table~5 of Ref.~\cite{US1}:
\begin{subequations}
\label{C6C8old}
\begin{eqnarray}
C_{6,V-A} &=& (-3\pm 4)\times 10^{-3}\ \mbox{GeV}^6 \qquad   (\mbox{FOPT})\ ,\label{C6C8olda}\\
          &=& (-4\pm 4)\times 10^{-3}\ \mbox{GeV}^6  \qquad  (\mbox{CIPT})\ ,\nonumber\\
C_{8,V-A} &=&  (-3\pm 12)\times 10^{-3}\ \mbox{GeV}^8 \qquad   (\mbox{FOPT})\ ,\label{C6C8oldb}\\
          &=&  (0\pm 12)\times 10^{-3}\ \mbox{GeV}^8  \qquad  (\mbox{CIPT})\ .\nonumber
\end{eqnarray}
\end{subequations}
The results for $C_{6,V-A}$ and $C_{8,V-A}$ are barely
consistent between the
updated and original OPAL data.   The relatively large differences between the ``updated'' and ``original''
data are not a big surprise:  these OPE coefficients parametrize the most
subleading part of the fits carried out in Refs.~\cite{US2,US1}.   Moreover,
it was found that the fits reported in Table~5 of Ref.~\cite{US2}, while consistent
with simpler fits, are at the ``statistical edge'' of what can be extracted
from the OPAL data.   
 
One can avoid using the fits of Table~5 of Ref.~\cite{US2} by employing the
sum rule~(\ref{cauchy}) with a judicious choice of the weights $w(t)$.
As we have seen, $\r_V(t)-\r_A(t)$ can be obtained from the simpler fits
reported in Table~3 of Ref.~\cite{US2}.   An obvious possibility is to choose
$w(t)=t^2$ or $w(t)=t^3$, for which the right-hand side of Eq.~(\ref{cauchy})
immediately yields $C_{6,V-A}$, respectively, $-C_{8,V-A}$.   We find
results consistent with those reported in Eq.~(\ref{C6C8}), with comparable
errors.

However, using the moments of Ref.~\cite{MY}, which involve a double-pinching
factor $(t-t_{\rm switch})^2$, we can do better.\footnote{This method was also
employed in Ref.~\cite{GPP3}.}   Choosing $w(t)=(t-t_{\rm switch})^2$ or
$w(t)=(t-t_{\rm switch})^2(t+2t_{\rm switch})$, Eq.~(\ref{cauchy}) implies
\begin{subequations}
\label{cauchy68}
\begin{eqnarray}
C_{6,V-A}&=&\sum_{i=1}^N\D t \;(t[i]-t_{\rm switch})^2\;
\left(\r_V(t[i])-\r_A(t[i])\right)-2f_\p^2(m_\p^2-t_{\rm switch})^2\nonumber\\
&&+\int_{t_{\rm switch}}^\infty dt\;(t-t_{\rm switch})^2\;\left(\r^{\rm DV}_V(t)-\r^{\rm DV}_A(t)\right)\ ,\label{cauchy68a}
\\
C_{8,V-A}&=&-\sum_{i=1}^N\D t \;(t[i]-t_{\rm switch})^2(t[i]+2t_{\rm switch})\;
\left(\r_V(t[i])-\r_A(t[i])\right)\nonumber\\
&&\qquad\ +2f_\p^2(m_\p^2-t_{\rm switch})^2(m_\p^2+2t_{\rm switch})\nonumber\\
&&-\int_{t_{\rm switch}}^\infty dt\;(t-t_{\rm switch})^2(t+2t_{\rm switch})\;\left(\r^{\rm DV}_V(t)-\r^{\rm DV}_A(t)\right)\ .\label{cauchy68b}
\end{eqnarray}
\end{subequations}
In these expressions, the sums over bins, as well as the pion-pole terms,
are obtained from data, the latter with negligible errors.\footnote{Order-$\a_s$ corrections from dimension two and four terms in the OPE
are again completely negligible.}   These sum rules have two advantages:
(1) they suppress the data at higher $t$, which have larger errors, and (2)
they suppress the contribution from the DV-integral terms \cite{MY,GPP3},
replacing these contributions, in effect, by the pion-pole terms, which are
known with great precision.

\begin{table}[t]
\begin{center}
\vspace*{2ex}
\begin{tabular}{|c|c|c|c|c|}
\hline
$s_{min}$  & $10^3C_{6,V-A}$ &  $10^3C^{\rm DV}_{6,V-A}$ & $10^3C_{8,V-A}$ &  $10^3C^{\rm DV}_{8,V-A}$ \\
\hline
1.408 &  $-7.3(5)$ &  0.8  & 8(2) & $-3$ \\
1.504 &  $-6.2(9)$ &  1.2  & 3(4) & $-5$ \\
1.600 &  $-6.4(8)$ &  0.3  & 4(4) & $-1$ \\
\hline
\hline
1.504 &  $-6.2(9)$ &  1.2  & 3(4) & $-5$ \\
\hline
\end{tabular}
\end{center}
\vspace*{4ex}
\caption{$C_{6,V-A}$ (in GeV$^6$) and $C_{8,V-A}$ (in GeV$^8$) from
Eq.~(\ref{cauchy68}).
The superscript DV indicates the part
coming from the DV integrals in Eq.~(\ref{cauchy68}).  Duality violation parameters are from the
fits of Ref.~\cite{US2}, Table~3.  Results from fits using FOPT are shown above the double line, those from CIPT below.}
\label{t3}
\end{table}%

We present the results in Table~\ref{t3}.   The DV parts are significantly smaller than those we would obtain with $w(t)=t^2$ or $w(t)=t^3$, especially for
$C_{6,V-A}$, but also for $C_{8,V-A}$.   And indeed, errors are also significantly smaller than those of Eq.~(\ref{C6C8}), as we expected.
We note, however, that there is some discrepancy between the values of
Table~\ref{t3} and Eq.~(\ref{C6C8}).   The results of Table~\ref{t3} are
based on results from simpler and more stable fits reported in 
Table~3 of Ref.~\cite{US2}.\footnote{For an extensive discussion of the quality of these fits, we refer to Ref.~\cite{US2}.}  Therefore, 
we take as our central results for $C_{6,V-A}$ and
$C_{8,V-A}$ the values from Table~\ref{t3} above, 
\begin{eqnarray}
\label{C6C8t3}
C_{6,V-A}&=&(-6.6\pm 1.1)\times 10^{-3}\ \mbox{GeV}^6
\ ,\\
C_{8,V-A}&=&(5\pm 5)\times 10^{-3}\ \mbox{GeV}^8
\ ,\nonumber
\end{eqnarray}
where the central values are the averages of the values in Table~\ref{t3},
and the errors have been obtained by adding the fitting error at $s_{min}=1.504$~GeV$^2$ and the variation as a function of $s_{min}$ in
quadrature.

In Fig.~\ref{f3} we compare our results with other results in the literature.
Updating
the OPAL data without including DVs in the analysis causes the
central values of the OPAL-based results of Ref.~\cite{CGM} to shift from $C_{6,V-A} = -5.4\times 10^{-3}$~GeV$^6$, $C_{8,V-A} = -1.4\times 10^{-3}$~GeV$^8$ to $C_{6,V-A} = -5.0\times 10^{-3}$~GeV$^6$ and $C_{8,V-A} = -3.4\times 10^{-3}$~GeV$^8$. A comparison of the latter set to the results of the present
analysis then shows directly the impact of the inclusion of DVs.\footnote{The reader should note that, for the $s_0$ employed in the fits of
Ref.~\cite{CGM}, integrated DVs have the opposite sign to those shown in 
Table~\ref{t3}.}

We note in particular that our values do not agree with those found in 
Ref.~\cite{GPP3}.   While our discussion above indicates that the determination
of  $C_{6,V-A}$ and $C_{8,V-A}$ is limited by the quality of the data,
we also recall that Ref.~\cite{GPP3} used a much more restricted parametrization
of duality violations in the $V-A$ channel, with four instead of eight 
parameters.   

\begin{figure}[t]
\centering
\includegraphics[scale=0.8]{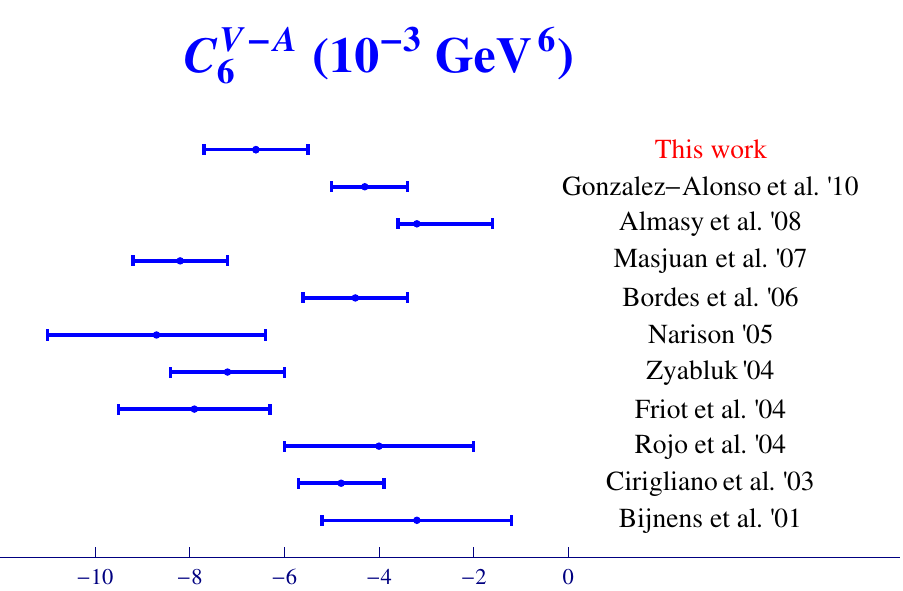}
\hspace{0.7cm}
\includegraphics[scale=0.8]{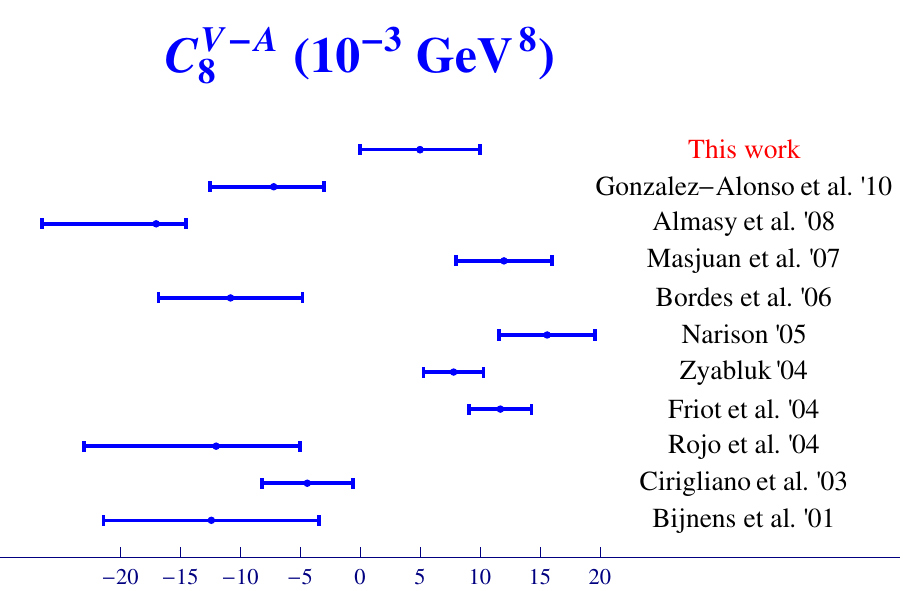}
\floatcaption{f3}{Comparison with other recent values \cite{GPP3,C6C8refs,CGM} for $C_{6,V-A}$ (left panel) and $C_{8,V-A}$ (right panel).}
\vspace*{2ex}
\end{figure}

It is interesting to compare our results for $C_{6,V-A}$ with an analytical
expression that is available at the next-to-leading order \cite{lsc86}
(see also ref.~\cite{bnp92}):
\begin{eqnarray}
\label{C6VmAvsa}
C_{6,V-A} &\!\!\!=\!\!\!& -\,\frac{32}{9}\,\pi\,\Big( 1 + \frac{119}{24\pi}\,
\alpha_s(s_0) \Big) \alpha_s(s_0) (\rho_1+\rho_5) \langle\bar qq(s_0)\rangle^2
 \\
\vbox{\vskip 8mm}
&& -\,\frac{2}{3}\,\alpha_s^2(s_0) (\tilde\rho_1+\tilde\rho_5)
\langle\bar qq(s_0)\rangle^2 \ .\nonumber
\end{eqnarray}
The parameters $\rho_{1,5}$ and $\tilde\rho_{1,5}$ parametrize deviations
from the so-called vacuum saturation approximation (VSA), in which
they are all normalized to unity. Values for $\rho_{1,5}$ from our fits have
already been discussed in Refs.~\cite{US2,US1}. Numerically, at
$s_0\approx m_\tau^2$ the second line of Eq.~(\ref{C6VmAvsa}) only contributes
about a few percent, so that precise values for $\tilde\r_{1,5}$ are irrelevant. On
the other hand, in the VSA the first line of Eq.~(\ref{C6VmAvsa}) yields
\begin{equation}
\label{C6VmAnum}
C_{6,V-A}^{\rm VSA} \,=\, -\,4.4\times 10^{-3}\,{\rm GeV}^6 \ ,
\end{equation}
where $\langle\bar qq(m_\t^2)\rangle=-(272\,{\rm MeV})^3$ \cite{bj08},
together with our result for $\a_s(m_\t^2)$ has been employed. As
the next-to-leading order correction in Eq.~(\ref{C6VmAvsa}) amounts to about
50\%, an error of that size should be attributed to the numerical value~(\ref{C6VmAnum}).
Therefore, the difference between our central fit result of Eqs.~(\ref{C6C8t3})
and~(\ref{C6VmAvsa}) could either be due to higher-order QCD corrections or
a breaking of the VSA. At any rate, no significant deviations from the VSA are
observed and the results in Eqs.~(\ref{C6C8t3}) and~(\ref{C6VmAnum})
are nicely compatible.

\vspace{0.8cm}
\section{\label{conclusion} Conclusion}
We used results of earlier fits to the non-strange vector- and axial-channel spectral
functions obtained from OPAL hadronic $\t$ decay data in order to estimate the low-energy constant
combinations $L_{10}^{\rm eff}$ and $C_{87}^{\rm eff}$, as well as the 
operator product coefficients $C_{6,V-A}$ and $C_{8,V-A}$.   Our best values are
\begin{eqnarray}
\label{final}
L_{10}^{\rm eff}&=&(-6.45\pm 0.09)\times 10^{-3}\ ,\\
C_{87}^{\rm eff}&=&(8.47\pm 0.29)\times 10^{-3}\ \mbox{GeV}^{-2}\ ,\nonumber\\
C_{6,V-A}&=&(-6.6\pm 1.1)\times 10^{-3}\ \mbox{GeV}^6\ ,\nonumber\\
C_{8,V-A}&=&(5\pm 5)\times 10^{-3}\ \mbox{GeV}^8\ .\nonumber
\end{eqnarray}
For a comparison with the values of $L_{10}^{\rm eff}$ and $C_{87}^{\rm eff}$
obtained in Ref.~\cite{GPP3}, we refer to Sec.~\ref{LECs}.
For comparisons with other values for $C_{6,V-A}$ and $C_{8,V-A}$ obtained
in the literature, see Fig.~\ref{f3}.   As emphasized in Sec.~\ref{OPE},
for $C_{6,V-A}$ and $C_{8,V-A}$ the results are rather sensitive to small
variations in the data, and to the details of the fits.   In contrast, we expect
the results for $L_{10}^{\rm eff}$ and $C_{87}^{\rm eff}$ to be rather robust,
since these values are dominated by the low-$Q^2$ range of the data,
where the experimental errors are small.   For a comparison of the low-$Q^2$
behavior of $\P_{V-A}(Q^2)$ with ChPT to order $p^6$, we refer to Sec.~\ref{ChPT}.   We find that order-$p^8$ effects, not included in our chiral
fits, are clearly visible in $C_{87}^r$, but not in $L_{10}^r$.   This is consistent
with what one would expect:  taking into account order-$p^6$ terms
stabilizes the values of the LECs at lower order.  
In Sec.~\ref{LECvalues} we presented and discussed preliminary estimates of $L_{10}^r$ and $C_{87}^r$.

We
demonstrated in both Ref.~\cite{US1} and Ref.~\cite{US2}, that our fits satisfy both Weinberg
sum rules, as well as the DGMLY sum rule 
for the pion
electromagnetic self-enegy \cite{EMpion} within errors, though none of
these were enforced in the fits.   
The situation is thus very much analogous
to that of the analysis of Refs.~\cite{GPP2} and \cite{GPP3}.  There, the set of
``acceptable'' DV parameter combinations  was generated by requiring
the corresponding DV contributions to the Weinberg and DGMLY
sum rules to be such that all three sum rules were satisfied
within the experimental errors on the data part of these sum rules,
\ie, the integral from 0 to $s_0$ in Eq.~(\ref{cauchy}).
On this point, there is thus no relevant difference between
the strategies employed in Refs.~\cite{US2,US1} and Refs.~\cite{GPP2,GPP3}.

There are important differences, however.
First, Refs.~\cite{GPP2,GPP3} started from an
{\it ansatz} of the form~(\ref{DVpar}) for the DV part of $\r_V-\r_A$ involving only four parameters rather than four for each of the two channels separately.  
The possibility
that the vector and axial DV contributions are such as to allow the $V-A$ combination
to be expressed in this simplified form, however, is not supported by
the results of our fits to the individual vector and axial channels.
Furthermore,  the procedure of Ref.~\cite{GPP2}, described in more detail in
Ref.~\cite{thesis}, does not
take into account the correlations between the data and DV parameters
induced by the use of the Weinberg and DGMLY sum rules.
Neither were the 
correlations between the data and the DV parameters taken into account when using their results
to evaluate the quantities of interest, $L_{10}^{\rm eff}$, \etc\ 
In our analysis, we have taken these correlations fully into account, and find
them to have a significant effect.

Finally, most of the earlier results shown in Fig.~\ref{f3} are based on ALEPH
data \cite{ALEPH,ALEPH98}.   At least for those
earlier works which employed the 2005/2008 version of these data
\cite{ALEPH}, the
incompleteness of the 2005/2008 correlation matrices \cite{TAU10} should be born in mind when appraising these results.
We wish to reiterate the expectation that inclusive spectral functions extracted
from BaBar or Belle would be of great help in reducing the uncertainties
on $C_{6,V-A}$ and $C_{8,V-A}$, for the reasons already discussed in
Ref.~\cite{US2}.

\vspace{3ex}
\noindent {\bf Acknowledgments}
\vspace{3ex}

KM thanks the
Department of Physics at the Universitat Aut\`onoma de Barcelona for hospitality.
DB is
supported by the Alexander von Humboldt Foundation,
MG is supported in part by the US Department of Energy, and in part by the Spanish Ministerio de Educaci\'on, Cultura y Deporte, under program SAB2011-0074.
MJ and SP are supported by CICYTFEDER-FPA2008-01430, FPA2011-25948, SGR2009-894,
the Spanish Consolider-Ingenio 2010 Program
CPAN (CSD2007-00042). 
KM is supported by a grant from the Natural Sciences and
Engineering Research Council of Canada.


\end{document}